\documentclass[longauth]{aa}  

\usepackage{graphicx}
\usepackage{txfonts}
\usepackage{lipsum}
\usepackage{xspace}
\usepackage{orcidlink}
\usepackage{makecell}
\usepackage{xcolor}

\hypersetup{
    colorlinks=true,
    allcolors=blue
}

\newcommand{\forestflow}{\texttt{ForestFlow}\xspace}

\newcommand{\mpgadget}{\texttt{MP-Gadget}\xspace}

\newcommand{\accel}{\texttt{ACCEL-2}\xspace}
\newcommand{\cupid}{\texttt{cup1d}\xspace}

\newcommand{\poned}{\ensuremath{P_{\rm 1D}}\xspace}
\newcommand{\pthreed}{\ensuremath{P_{\rm 3D}}\xspace}
\newcommand{\lya}{\ensuremath{\mathrm{Ly}\alpha}\xspace}
\newcommand{\lyaf}{\ensuremath{\mathrm{Ly}\alpha~ \mathrm{forest}}\xspace}

\newcommand{\Mpc}{\ensuremath{\mathrm{Mpc}}}
\newcommand{\hMpc}{\ensuremath{h^{-1}\mathrm{Mpc}}}

\newcommand{\iMpc}{\ensuremath{\mathrm{Mpc}^{-1}}}

\newcommand{\kms}{\ensuremath{\mathrm{km}\,\mathrm{s}^{-1}}\xspace}
\newcommand{\zeff}{\ensuremath{z_\mathrm{eff}}}

\begin{document}

\title{Lyman-$\alpha$ forest holography: 3D predictions from 1D measurements}
\titlerunning{Lyman-$\alpha$ forest holography}

\authorrunning{J.~Chaves-Montero et al.}

\author{
J.~Chaves-Montero\orcidlink{0000-0002-9553-4261} \inst{1},
A.~Font-Ribera\orcidlink{0000-0002-3033-7312} \inst{1,2},
J.~Aguilar \inst{3},
S.~Ahlen\orcidlink{0000-0001-6098-7247} \inst{4},
E.~Armengaud\orcidlink{0000-0001-7600-5148} \inst{5},
A.~Aviles\orcidlink{0000-0001-5998-3986} \inst{6,7,8},
F.~Beutler\orcidlink{0000-0003-0467-5438} \inst{9},
D.~Bianchi\orcidlink{0000-0001-9712-0006} \inst{10,11},
S.~Blasby \inst{12},
D.~Brooks \inst{13},
K.~Carrion\orcidlink{0000-0002-1798-7978} \inst{8},
Z.~Chen  \inst{9},
T.~Claybaugh \inst{3},
A.~Cuceu\orcidlink{0000-0002-2169-0595} \inst{3},
A.~de la Macorra\orcidlink{0000-0002-1769-1640} \inst{8},
A.~Dey\orcidlink{0000-0002-4928-4003} \inst{14},
P.~Doel \inst{13},
W.~Elbers\orcidlink{0000-0002-2207-6108} \inst{15},
S.~Ferraro\orcidlink{0000-0003-4992-7854} \inst{3,16},
L.~Flores\orcidlink{0009-0009-9264-7410} \inst{1},
J.~E.~Forero-Romero\orcidlink{0000-0002-2890-3725} \inst{17,18},
E.~Gaztañaga\orcidlink{0000-0001-9632-0815} \inst{19,20,21},
S.~Gontcho A Gontcho\orcidlink{0000-0003-3142-233X} \inst{22},
D.~Gonzalez\orcidlink{0009-0009-6485-640X} \inst{23},
A.~X.~Gonzalez-Morales\orcidlink{0000-0003-4089-6924} \inst{23},
R.~Gsponer\orcidlink{0000-0002-7540-7601} \inst{24},
G.~Gutierrez \inst{25},
C.~Hahn\orcidlink{0000-0003-1197-0902} \inst{26},
M.~Herbold\orcidlink{0009-0000-8112-765X} \inst{27},
H.~K.~Herrera-Alcantar\orcidlink{0000-0002-9136-9609} \inst{5,28},
K.~Honscheid\orcidlink{0000-0002-6550-2023} \inst{27,29,30},
M.~Ishak\orcidlink{0000-0002-6024-466X} \inst{31},
S.~Juneau\orcidlink{0000-0002-0000-2394} \inst{14},
N.~V.~Kamble\orcidlink{0009-0008-6707-2777} \inst{31},
D.~Kirkby\orcidlink{0000-0002-8828-5463} \inst{32},
A.~Kremin\orcidlink{0000-0001-6356-7424} \inst{3},
A.~Lambert \inst{3},
M.~Landriau\orcidlink{0000-0003-1838-8528} \inst{3},
L.~Le~Guillou\orcidlink{0000-0001-7178-8868} \inst{33},
K.~Lodha\orcidlink{0009-0004-2558-5655} \inst{34,35},
Z.~Luki\'c \inst{3},
M.~Manera\orcidlink{0000-0003-4962-8934} \inst{1,36},
P.~Martini\orcidlink{0000-0002-4279-4182} \inst{27,29,37},
A.~Meisner\orcidlink{0000-0002-1125-7384} \inst{14},
R.~Miquel \inst{1,2},
P.~Mukherjee\orcidlink{0000-0002-2701-5654} \inst{34},
A.~Muñoz-Gutiérrez \inst{8},
S.~Nadathur\orcidlink{0000-0001-9070-3102} \inst{20},
H.~E.~Noriega\orcidlink{0000-0002-3397-3998} \inst{7,8},
E.~Paillas\orcidlink{0000-0002-4637-2868} \inst{38,39},
N.~Palanque-Delabrouille\orcidlink{0000-0003-3188-784X} \inst{3,5},
W.~J.~Percival\orcidlink{0000-0002-0644-5727} \inst{12,40,41},
C.~Poppett \inst{3,16,42},
F.~Prada\orcidlink{0000-0001-7145-8674} \inst{43},
H.~Pulido-Hern{\'a}ndez\orcidlink{0009-0009-7807-9218} \inst{44},
I.~P\'erez-R\`afols\orcidlink{0000-0001-6979-0125} \inst{45},
C.~Ravoux\orcidlink{0000-0002-3500-6635} \inst{46},
J.~Rohlf\orcidlink{0000-0001-6423-9799} \inst{4},
A.~J.~Rosado-Mar\'{i}n\orcidlink{0000-0001-7545-3504} \inst{47},
G.~Rossi \inst{48},
R.~Ruggeri\orcidlink{0000-0002-0394-0896} \inst{49},
M. F.~Ruiz-Herrera Bernal\orcidlink{0009-0000-5572-6157} \inst{50},
E.~Sanchez\orcidlink{0000-0002-9646-8198} \inst{50},
C.~Saulder\orcidlink{0000-0002-0408-5633} \inst{51},
D.~Schlegel \inst{3},
M.~Schubnell \inst{52,53},
F.~Sinigaglia\orcidlink{0000-0002-0639-8043} \inst{54,55},
G.~Tarl\'{e}\orcidlink{0000-0003-1704-0781} \inst{53},
W.~Turner\orcidlink{0009-0008-3418-5599} \inst{27,29,37},
B.~A.~Weaver \inst{14},
and H.~Zhang\orcidlink{0000-0001-6847-5254} \inst{12,41}
}

\institute{
Institut de F\'{i}sica d’Altes Energies (IFAE), The Barcelona Institute of Science and Technology, Edifici Cn, Campus UAB, 08193, Bellaterra (Barcelona), Spain 
\and Instituci\'{o} Catalana de Recerca i Estudis Avan\c{c}ats, Passeig de Llu\'{\i}s Companys, 23, 08010 Barcelona, Spain 
\and Lawrence Berkeley National Laboratory, 1 Cyclotron Road, Berkeley, CA 94720, USA 
\and Department of Physics, Boston University, 590 Commonwealth Avenue, Boston, MA 02215 USA 
\and IRFU, CEA, Universit\'{e} Paris-Saclay, F-91191 Gif-sur-Yvette, France 
\and Instituto Avanzado de Cosmolog\'{\i}a A.~C., San Marcos 11 - Atenas 202. Magdalena Contreras. Ciudad de M\'{e}xico C.~P.~10720, M\'{e}xico 
\and Instituto de Ciencias F\'{\i}sicas, Universidad Nacional Aut\'onoma de M\'exico, Av. Universidad s/n, Cuernavaca, Morelos, C.~P.~62210, M\'exico 
\and Instituto de F\'{\i}sica, Universidad Nacional Aut\'{o}noma de M\'{e}xico,  Circuito de la Investigaci\'{o}n Cient\'{\i}fica, Ciudad Universitaria, Cd. de M\'{e}xico  C.~P.~04510,  M\'{e}xico 
\and Institute for Astronomy, University of Edinburgh, Royal Observatory, Blackford Hill, Edinburgh EH9 3HJ, UK 
\and Dipartimento di Fisica ``Aldo Pontremoli'', Universit\`a degli Studi di Milano, Via Celoria 16, I-20133 Milano, Italy 
\and INAF-Osservatorio Astronomico di Brera, Via Brera 28, 20122 Milano, Italy 
\and Department of Physics and Astronomy, University of Waterloo, 200 University Ave W, Waterloo, ON N2L 3G1, Canada 
\and Department of Physics \& Astronomy, University College London, Gower Street, London, WC1E 6BT, UK 
\and NSF NOIRLab, 950 N. Cherry Ave., Tucson, AZ 85719, USA 
\and Institute for Computational Cosmology, Department of Physics, Durham University, South Road, Durham DH1 3LE, UK 
\and University of California, Berkeley, 110 Sproul Hall \#5800 Berkeley, CA 94720, USA 
\and Departamento de F\'isica, Universidad de los Andes, Cra. 1 No. 18A-10, Edificio Ip, CP 111711, Bogot\'a, Colombia 
\and Observatorio Astron\'omico, Universidad de los Andes, Cra. 1 No. 18A-10, Edificio H, CP 111711 Bogot\'a, Colombia 
\and Institut d'Estudis Espacials de Catalunya (IEEC), c/ Esteve Terradas 1, Edifici RDIT, Campus PMT-UPC, 08860 Castelldefels, Spain 
\and Institute of Cosmology and Gravitation, University of Portsmouth, Dennis Sciama Building, Portsmouth, PO1 3FX, UK 
\and Institute of Space Sciences, ICE-CSIC, Campus UAB, Carrer de Can Magrans s/n, 08913 Bellaterra, Barcelona, Spain 
\and University of Virginia, Department of Astronomy, Charlottesville, VA 22904, USA 
\and Departamento de F\'{\i}sica, DCI-Campus Le\'{o}n, Universidad de Guanajuato, Loma del Bosque 103, Le\'{o}n, Guanajuato C.~P.~37150, M\'{e}xico 
\and Institute of Physics, Laboratory of Astrophysics, \'{E}cole Polytechnique F\'{e}d\'{e}rale de Lausanne (EPFL), Observatoire de Sauverny, Chemin Pegasi 51, CH-1290 Versoix, Switzerland 
\and Fermi National Accelerator Laboratory, PO Box 500, Batavia, IL 60510, USA 
\and Department of Astronomy, University of Texas at Austin, 2515 Speedway, TX 78712, USA 
\and The Ohio State University, Columbus, 43210 OH, USA 
\and Institut d'Astrophysique de Paris. 98 bis boulevard Arago. 75014 Paris, France 
\and Center for Cosmology and AstroParticle Physics, The Ohio State University, 191 West Woodruff Avenue, Columbus, OH 43210, USA 
\and Department of Physics, The Ohio State University, 191 West Woodruff Avenue, Columbus, OH 43210, USA 
\and Department of Physics, The University of Texas at Dallas, 800 W. Campbell Rd., Richardson, TX 75080, USA 
\and Department of Physics and Astronomy, University of California, Irvine, 92697, USA 
\and Sorbonne Universit\'{e}, CNRS/IN2P3, Laboratoire de Physique Nucl\'{e}aire et de Hautes Energies (LPNHE), FR-75005 Paris, France 
\and Korea Astronomy and Space Science Institute, 776, Daedeokdae-ro, Yuseong-gu, Daejeon 34055, Republic of Korea 
\and University of Science and Technology, 217 Gajeong-ro, Yuseong-gu, Daejeon 34113, Republic of Korea 
\and Departament de F\'{i}sica, Serra H\'{u}nter, Universitat Aut\`{o}noma de Barcelona, 08193 Bellaterra (Barcelona), Spain 
\and Department of Astronomy, The Ohio State University, 4055 McPherson Laboratory, 140 W 18th Avenue, Columbus, OH 43210, USA 
\and Instituto de Estudios Astrof\'isicos, Facultad de Ingenier\'ia y Ciencias, Universidad Diego Portales, Av. Ej\'ercito Libertador 441, Santiago, Chile 
\and Steward Observatory, University of Arizona, 933 N. Cherry Avenue, Tucson, AZ 85721, USA 
\and Perimeter Institute for Theoretical Physics, 31 Caroline St. North, Waterloo, ON N2L 2Y5, Canada 
\and Waterloo Centre for Astrophysics, University of Waterloo, 200 University Ave W, Waterloo, ON N2L 3G1, Canada 
\and Space Sciences Laboratory, University of California, Berkeley, 7 Gauss Way, Berkeley, CA  94720, USA 
\and Instituto de Astrof\'{i}sica de Andaluc\'{i}a (CSIC), Glorieta de la Astronom\'{i}a, s/n, E-18008 Granada, Spain 
\and Departamento de F\'{i}sica, Instituto Nacional de Investigaciones Nucleares, Carreterra M\'{e}xico-Toluca S/N, La Marquesa,  Ocoyoacac, Edo. de M\'{e}xico C.~P.~52750,  M\'{e}xico 
\and Departament de F\'isica, EEBE, Universitat Polit\`ecnica de Catalunya, c/Eduard Maristany 10, 08930 Barcelona, Spain 
\and Universit\'{e} Clermont-Auvergne, CNRS, LPCA, 63000 Clermont-Ferrand, France 
\and Department of Physics \& Astronomy, Ohio University, 139 University Terrace, Athens, OH 45701, USA 
\and Department of Physics and Astronomy, Sejong University, 209 Neungdong-ro, Gwangjin-gu, Seoul 05006, Republic of Korea 
\and Queensland University of Technology,  School of Chemistry \& Physics, George St, Brisbane 4001, Australia 
\and CIEMAT, Avenida Complutense 40, E-28040 Madrid, Spain 
\and Max Planck Institute for Extraterrestrial Physics, Gie\ss enbachstra\ss e 1, 85748 Garching, Germany 
\and Department of Physics, University of Michigan, 450 Church Street, Ann Arbor, MI 48109, USA 
\and University of Michigan, 500 S. State Street, Ann Arbor, MI 48109, USA 
\and Departamento de Astrof\'{\i}sica, Universidad de La Laguna (ULL), E-38206, La Laguna, Tenerife, Spain 
\and Instituto de Astrof\'{\i}sica de Canarias, C/ V\'{\i}a L\'{a}ctea, s/n, E-38205 La Laguna, Tenerife, Spain 
}

\abstract
{
Cosmological analyses of Lyman-$\alpha$ forest clustering rely on either one-dimensional correlations along individual sightlines or three-dimensional correlations between different sightlines. Because these observables probe the matter distribution on very different scales, they have traditionally been analyzed independently. In this work, we bridge this gap using \texttt{ForestFlow}, an emulator trained on a suite of cosmological hydrodynamical simulations that provides a unified description of Lyman-$\alpha$ forest clustering from linear to nonlinear scales. This framework enables us to determine the range of three-dimensional clustering models compatible with the DESI one-dimensional flux power spectrum ($P_{\rm 1D}$). The resulting predictions successfully reproduce the large-scale clustering measured by the DESI BAO analysis and provide physically motivated priors on nonlinear clustering that are used in a companion paper presenting the full-shape analysis of the DESI DR2 Lyman-$\alpha$ forest. We validate our methodology using the large-volume, high-resolution hydrodynamical simulation \texttt{ACCEL-2}, demonstrating excellent agreement across the full range of scales considered. Finally, we combine constraints from the $P_{\rm 1D}$ and BAO analyses on the parameter combinations $b_\delta \sigma_8$ and $b_\eta f \sigma_8$, finding that the two probes provide comparable constraining power while exhibiting complementary parameter degeneracies. Our results establish a direct connection between one- and three-dimensional Lyman-$\alpha$ forest measurements through \texttt{ForestFlow}, an approach we term \emph{Lyman-$\alpha$ holography} by analogy with the reconstruction of higher-dimensional structure from lower-dimensional information.}

\keywords{cosmology: observations; cosmology: large-scale structure of Universe; intergalactic medium; quasars: absorption lines}

\maketitle

\section{Introduction} 
\label{sec:intro}

The Lyman-$\alpha$ (\lya) forest consists of absorption features in the spectra of distant sources produced by intervening neutral hydrogen in the intergalactic medium \citep[see][for comprehensive reviews]{Rauch1998, Weinberg2003, Meiksin2009, McQuinn2016, ChavesMontero2026_ML}. Unlike galaxies and quasars, which trace the matter distribution at discrete locations, the absorption features in a single spectrum probe the matter distribution continuously along the line of sight over a broad range of redshifts. As a result, analyses of correlations along individual sightlines, commonly characterized by the one-dimensional \lya flux power spectrum (\poned), provide precise measurements of the linear matter power spectrum on small scales \citep{McDonald2006, PalanqueDelabrouille2015, Chabanier2019, Fernandez2024, Walther2025, ChavesMontero2026}. Conversely, three-dimensional correlations between different sightlines have enabled some of the most precise measurements of baryon acoustic oscillations \citep[BAO;][]{Busca2013, Slosar2013, Delubac2015, Bautista2017, dSA2019, dMdB2020, DESI2024.IV.KP6, DESI.DR2.BAO.lya}.

The statistical power of \lya forest studies has steadily increased with the growing number of quasar spectra provided by spectroscopic surveys. For nearly two decades, the successive phases of the Sloan Digital Sky Survey \citep[SDSS;][]{york2000_sdss, Dawson2013, Dawson2016} delivered the largest \lya forest samples available, measuring 210\,000 forests in its sixteenth data release. This situation changed with the advent of the Dark Energy Spectroscopic Instrument \citep[DESI;][]{DESI2016a.Science}, which has dramatically expanded the dataset, measuring approximately 450\,000 forests in its first data release \citep[DR1;][]{DESI2024.I.DR1} and 820\,000 in its second \citep[DR2;][]{DESI.DR2.DR2}.

Despite probing the same underlying matter distribution, \poned and BAO analyses have traditionally been carried out independently because they extract cosmological information from distinct physical regimes. BAO analyses rely on large, linear scales that can be modeled accurately using perturbation theory, whereas \poned analyses derive much of their constraining power from highly nonlinear scales, whose interpretation requires cosmological hydrodynamical simulations. As a consequence, the cosmological information encoded on small and large scales has not yet been exploited jointly within a unified inference framework.

To address this challenge, \citet{ChavesMontero2025} introduced \forestflow\footnote{\url{https://github.com/igmhub/ForestFlow}}, an emulator trained on a suite of cosmological hydrodynamical simulations that provides a unified description of \lya forest clustering from linear to nonlinear scales. In this work, we use this model to predict the range of three-dimensional clustering models consistent with constraints from the DESI DR1 \poned analysis \citep{ChavesMontero2026}. We refer to this approach as ``holography'', in analogy with the way a hologram encodes and reconstructs higher-dimensional structure from lower-dimensional projections. The resulting model-dependent predictions reproduce the large-scale clustering measured by the DESI DR1 and DR2 BAO analyses \citep{DESI2024.IV.KP6, DESI.DR2.BAO.lya} and provide physically motivated priors on small-scale clustering that are used in a companion work presenting the full-shape analysis of DESI DR2 \lya measurements \citep{DESIDR2full}.

The remainder of this paper is organized as follows. In Section~\ref{sec:data}, we briefly describe the DESI \poned and BAO measurements. In Section~\ref{sec:mapping}, we present our methodology for inferring the range of three-dimensional clustering models compatible with one-dimensional measurements, and we validate it using a high-resolution hydrodynamical simulation. In Section~\ref{sec:combined}, we compare and combine constraints from one- and three-dimensional analyses, highlighting their complementarity. Finally, in Section~\ref{sec:conclusions}, we summarize our main results and discuss their implications for a self-consistent joint analyses of \lya forest clustering.
\section{DESI data}
\label{sec:data}

DESI is a multi-fiber spectroscopic instrument mounted on the Mayall 4-meter telescope at Kitt Peak National Observatory \citep{DESI2016b.Instr, DESI2022.KP1.Instr}. Its focal plane \citep{FocalPlane.Silber.2023}, corrector \citep{Corrector.Miller.2023}, fiber-positioning system \citep{FiberSystem.Poppett.2024}, and survey operations strategy \citep{SurveyOps.Schlafly.2023} enable the simultaneous observation of nearly 5000 targets per exposure. The DESI spectrographs cover the wavelength range $3600-9800\,\AA$ at a spectral resolution of $R\sim2000-5000$ \citep{Spectro.Pipeline.Guy.2023}, enabling precise measurements of the Lyman-$\alpha$ forest in quasars at $z \gtrsim 2.1$. In DESI, we identify these quasars and determines their redshifts through a multi-stage classification pipeline \citep{Busca2018, LS.Overview.Dey.2019, QSOPrelim.Yeche.2020, Farr20_QN, QSO.TS.Chaussidon.2023, VIQSO.Alexander.2023, Napolitano2023_mgii, KP6s4-Bault, RedrockQSO.Brodzeller.2023}.

From the resulting spectra, the DESI pipeline extracts the \lya forests measurements used in BAO and \poned analyses. Before doing so, it removes or mitigates known contaminants that are not associated with intergalactic hydrogen. In particular, the pipeline masks damped Lyman-$\alpha$ systems (DLAs), which produce strong, wide absorption features that can significantly bias clustering measurements \citep{Ho2021, Wang2022, Y3.lya-s2.Brodzeller.2025}, broad absorption line (BAL) quasars associated with quasar outflows \citep{KP6s9-Martini, Filbert2024_bal}, and regions affected by strong atmospheric emission lines and Galactic absorption \citep{2023MNRAS.tmp.3626R}.

The DESI pipeline extracts the \lya flux overdensity field from the cleaned spectra as follows
\begin{equation}
    \label{eq:deltas}
    \delta_q(\lambda) = \frac{f_q(\lambda)}{\bar{F}(z)C_q(\lambda)} -1,
\end{equation}
where $f_q$ is the observed flux density of quasar $q$, $C_q$ is the intrinsic (unabsorbed) quasar continuum, and $\bar{F}(z)$ is the mean transmitted flux fraction at redshift $z = \lambda/\lambda_\alpha - 1$, with $\lambda_\alpha$ the rest-frame \lya wavelength. Estimating the denominator constitutes the continuum-fitting procedure, which is a key step in all \lyaf analyses \citep{dMdB2020, 2023MNRAS.tmp.3626R}. Finally, the one- and three-dimensional clustering is measured from the flux overdensity field.

%%%%%%%%%%%%%%%%%%%%%%%%%%%%
%%%%%%%%%%%%%%%%%%%%%%%%%%%%

\subsection{\poned measurements}
\label{sec:data_p1d}

DESI \poned measurements are obtained using two independent methods: a quadratic maximum-likelihood estimator \citep[QMLE;][]{McDonald2006, Karacaily2020, Karacayli2024_edr, Karacayli2025_p1d_dr1} and Fast Fourier Transforms \citep[FFTs;][]{Ravoux2023, ravoux2025_p1d_dr1}. Their interpretation requires cosmological hydrodynamical simulations, as these measurements probe nonlinear scales down to $k \simeq 4\,\iMpc$ at high redshift. However, exploring the relevant cosmological and astrophysical parameter space with direct simulations is computationally prohibitive. As a result, cosmological analyses typically rely on surrogate models, or emulators, which interpolate simulation outputs as a function of the underlying parameters. 

In the baseline DESI DR1 \poned analysis, the \poned measurements obtained with the QMLE estimator \citep[][]{Karacayli2025_p1d_dr1} were modeled using \texttt{lace-mpg}\footnote{\url{https://github.com/igmhub/lace}} \citep{ChavesMontero2026}, an emulator that predicts \poned as a function of cosmology and the physical properties of the intergalactic medium (IGM). The emulator predictions were supplemented with analytical corrections to account for contaminants and systematic effects, and were then compared with the data using the likelihood code \cupid\footnote{\url{https://github.com/igmhub/cup1d}} \citep{Pedersen2021, Pedersen2023, ChavesMontero2026} to constrain cosmology and IGM physics. In Section~\ref{sec:mapping}, we use these constraints as input to \forestflow to infer the range of three-dimensional clustering models consistent with the DESI DR1 \poned measurements.

The \texttt{lace-mpg} emulator was trained on a suite of 30 fixed-and-paired hydrodynamical simulations \citep{Pedersen2021, CabayolGarcia2023} performed with the \texttt{TreeSPH} code \mpgadget \citep{feng2018MpGadgetMpGadgetTag}. It models the dependence of \poned on cosmology and IGM physics using six parameters: the amplitude and slope of the linear matter power spectrum at the pivot scale $k_\mathrm{p}=0.7\,\iMpc$, $\Delta^2_\mathrm{p}(z)$ and $n_\mathrm{p}(z)$; the mean transmitted flux fraction, $\bar{F}(z)$; the pressure-smoothing scale, $k_\mathrm{F}(z)$; and the amplitude and slope of the temperature--density relation,
\begin{equation}
T(z) = T_0(z)\Delta_b^{\gamma(z)-1},
\end{equation}
where $\Delta_b$ is the baryon overdensity. Rather than using $T_0(z)$ directly, the emulator takes as input the amplitude of the temperature--density relation expressed in comoving velocity units, $\sigma_\mathrm{T}(z)$, which is related to $T_0(z)$ through
\begin{equation}
\label{eq:t0}
T_0(z) = \left[\frac{\sigma_\mathrm{T}(z)}{\sigma_\mathrm{T,0}} \frac{H(z)}{1+z} \right]^2 10^4\,\mathrm{K},
\end{equation}
where $\sigma_\mathrm{T,0} = 9.1\,\kms$ and $H(z)$ is the Hubble parameter at redshift $z$.

Using this framework, \citet{ChavesMontero2026} constrained the amplitude and slope of the linear matter power spectrum at $z=3$, together with a set of parameters describing the redshift evolution of the IGM. To translate these constraints into the parameterization used by \texttt{lace-mpg}, we randomly select 10\,000 samples from the Markov Chain Monte Carlo (MCMC) chain of the DESI \poned analysis. For each sample, we evaluate \cupid to compute the values of the emulator input parameters at the redshifts probed by the \poned measurements, as well as at the effective redshift of the BAO analysis, $z_\mathrm{eff}=2.33$ (see the next section). The resulting mean values and standard deviations are listed in Table~\ref{tab:poned_data}.

%%%%%%%%%%%%%%%%%%%%%%%%%%%%
\begin{table*}[]
    \caption{Constraints on cosmology and IGM physics from the analysis of DESI \poned measurements \citep{ChavesMontero2026}. Not included in the table, the best-fitting value of the power spectrum slope is $n_\mathrm{p}=-2.313 \pm 0.020$ for all redshifts. See Eq.~\ref{eq:t0} for the relation between $\sigma_\mathrm{T}$ and $T_0$.}
    \label{tab:poned_data}
    \centering
    \begin{tabular}{c|ccccc}
        $z$ & $\Delta^2_\mathrm{p}$ & $\bar{F}$ & $\sigma_\mathrm{T}\,[\Mpc]$ & $\gamma$ & $k_\mathrm{F}\,[\iMpc]$\\
        \hline
        2.20 & $0.592 \pm 0.052$ & $0.8151 \pm 0.0047$ & $0.1611 \pm 0.0043$ & $1.859 \pm 0.084$ & $8.36 \pm 0.50$ \\
        2.33 & $0.548 \pm 0.048$ & $0.7994 \pm 0.0044$ & $0.1519 \pm 0.0036$ & $1.740 \pm 0.060$ & $8.49 \pm 0.41$ \\
        2.40 & $0.526 \pm 0.046$ & $0.7906 \pm 0.0044$ & $0.1467 \pm 0.0037$ & $1.677 \pm 0.053$ & $8.57 \pm 0.38$ \\
        2.60 & $0.471 \pm 0.041$ & $0.7637 \pm 0.0049$ & $0.1311 \pm 0.0057$ & $1.497 \pm 0.067$ & $8.81 \pm 0.44$ \\
        2.80 & $0.424 \pm 0.037$ & $0.7303 \pm 0.0056$ & $0.1205 \pm 0.0068$ & $1.430 \pm 0.082$ & $9.12 \pm 0.52$ \\
        3.00 & $0.383 \pm 0.034$ & $0.6850 \pm 0.0054$ & $0.1196 \pm 0.0059$ & $1.566 \pm 0.056$ & $9.54 \pm 0.51$ \\
        3.20 & $0.348 \pm 0.030$ & $0.6349 \pm 0.0061$ & $0.1181 \pm 0.0069$ & $1.700 \pm 0.042$ & $10.00 \pm 0.65$ \\
        3.40 & $0.318 \pm 0.028$ & $0.5808 \pm 0.0076$ & $0.1160 \pm 0.0088$ & $1.812 \pm 0.050$ & $10.55 \pm 0.85$ \\
        3.60 & $0.291 \pm 0.025$ & $0.5319 \pm 0.0065$ & $0.1141 \pm 0.0070$ & $1.660 \pm 0.042$ & $11.63 \pm 0.76$ \\
        3.80 & $0.268 \pm 0.023$ & $0.4810 \pm 0.0056$ & $0.1116 \pm 0.0064$ & $1.510 \pm 0.050$ & $12.81 \pm 0.87$ \\
        4.00 & $0.247 \pm 0.022$ & $0.4288 \pm 0.0057$ & $0.1085 \pm 0.0072$ & $1.362 \pm 0.068$ & $14.1 \pm 1.2$ \\
        4.20 & $0.228 \pm 0.020$ & $0.3762 \pm 0.0067$ & $0.1048 \pm 0.0086$ & $1.216 \pm 0.090$ & $15.5 \pm 1.6$ \\
    \end{tabular}
\end{table*}
%%%%%%%%%%%%%%%%%%%%%%%%%%%%

%%%%%%%%%%%%%%%%%%%%%%%%%%%%
%%%%%%%%%%%%%%%%%%%%%%%%%%%%

\subsection{BAO measurements}
\label{sec:data_xi3d}

In DESI, we extract BAO information from the three-dimensional auto-correlation of the \lya flux overdensity and its cross-correlation with quasar positions \citep{2023JCAP...11..045G, DESI2024.IV.KP6, DESI.DR2.BAO.lya}. Together with galaxy and quasar BAO measurements, these provide some of the tightest constraints on the expansion history of the Universe \citep{DESI2024.VI.KP7A, DESI2024.VII.KP7B, DESI.DR2.BAO.cosmo}. The correlation functions and covariance matrices of \lya clustering are computed using the \texttt{picca}\footnote{\url{https://github.com/igmhub/picca}} package \citep{picca}, while the modeling of the correlations is performed with the \texttt{Vega}\footnote{\url{https://github.com/andreicuceu/vega}} package. We refer the reader to \citet{2023JCAP...11..045G, KP6s6-Cuceu, Y3.lya-s1.Casas.2025} for a detailed description of the methodology and summarize here only the aspects most relevant to the present work.

The correlations are measured in bins of comoving separation parallel and perpendicular to the line of sight, where angular and redshift separations are converted into distances assuming a fiducial cosmology corresponding to the best-fitting $\Lambda$CDM model from \textit{Planck} 2018 \citep{Planck2018}. The analysis is performed in a single redshift bin containing all \lya-\lya and \lya-quasar pairs, with an effective redshift of $z_\mathrm{eff}=2.33$. The clustering model is based on the linear matter power spectrum at this redshift and includes the effects of large-scale redshift-space distortions \citep{Kaiser1987, McDonald2003}. In this way, the \lya auto-correlation depends on the large-scale bias parameters $b_\delta$ and $\beta \equiv b_\eta f/b_\delta$, while the \lya-quasar cross-correlation additionally depends on the quasar bias parameter $b_\mathrm{Q}$. The full model further accounts for several astrophysical and observational effects, including metal absorption, high column density absorbers (HCDs), quasar redshift uncertainties, and the distortion matrix.

Even though the primary goal of BAO analyses is to measure the dilation parameters $\alpha_\parallel$ and $\alpha_\perp$, the overall large-scale clustering amplitude and its anisotropy are sensitive to the combinations $b_\delta\,\sigma_8$ and $\beta$, respectively. In standard BAO analyses, however, the template cosmology is held fixed and only $b_\delta$ and $\beta$ are varied during the fit. Consequently, the inferred value of $b_\delta$ should be interpreted as a measurement of $b_\delta\,\sigma_8/\sigma_8^\mathrm{fid}$, where $\sigma_8^\mathrm{fid}=0.307$ denotes the amplitude of matter fluctuations in the fiducial cosmology at the effective redshift of the measurement.

These quantities are particularly relevant for the present work because they can be directly compared with the predictions obtained by propagating \poned constraints through \forestflow. However, modeling choices that have a negligible impact on the BAO parameters can significantly affect the inferred values of $b_\delta\,\sigma_8/\sigma_8^\mathrm{fid}$ and $\beta$. For example, the range of scales included in the fit has little effect on $\alpha_\parallel$ and $\alpha_\perp$, but can alter the inferred level of nuisance contributions such as HCD contamination, which is strongly degenerate with the \lya bias parameters.

As explained in Appendix~\ref{app:hcd_prior}, we reanalyze the DESI DR1 and DR2 BAO measurements \citep{DESI2024.IV.KP6, DESI.DR2.BAO.lya} to enable a consistent comparison of the inferred large-scale \lya bias parameters. Specifically, we adopt the same fitting range, restrict the analysis to a high signal-to-noise subsample for which the HCD contamination is reduced, and impose a prior on the amplitude of this contamination. Table~\ref{tab:BAO_data} summarizes the resulting constraints on $b_\delta\,\sigma_8/\sigma_8^\mathrm{fid}$ and $\beta$, together with those on the parameter combinations $b_\delta \sigma_8$ and $b_\eta f \sigma_8$, obtained without applying the $\sigma_8^\mathrm{fid}$ normalization.

%%%%%%%%%%%%%%%%%%%%%%%%%%%%
\begin{table}
\caption{
Best-fitting values of the large-scale \lya bias parameters obtained from our reanalyses of the DESI BAO measurements (see Appendix~\ref{app:hcd_prior}). The first set of rows reports the constraints on the large-scale \lya biases when assuming the \textit{Planck} 2018 value for the amplitude of matter fluctuations at the effective redshift, $\sigma_8^\mathrm{fid}(z_\mathrm{eff}=2.33)=0.307$. The second set reports the corresponding constraints without assuming the previous value.}
\label{tab:BAO_data}
\centering
\begin{tabular}{c|cc}
Parameter          &   DESI DR1                       & DESI DR2  \\
\hline
$b_\delta\, \sigma_8/ \sigma_8^\mathrm{fid}$ & $-0.1316 \pm 0.0058$ & $-0.1301 \pm 0.0048$ \\
$\beta$   & $1.546 \pm 0.094$   & $1.502 \pm 0.066$ \\
\hline
$b_\delta\, \sigma_8$  & $-0.0404 \pm 0.0018$ & $-0.0399 \pm 0.0015$ \\
$b_\eta f \sigma_8$  & $-0.0622 \pm 0.0018$ & $-0.0599 \pm 0.0013$ \\
\end{tabular}
\end{table}
%%%%%%%%%%%%%%%%%%%%%%%%%%%%

\section{Connecting one- and three-dimensional analyses}
\label{sec:mapping}

BAO and \poned analyses have traditionally been performed independently because they probe different physical regimes. However, the recent development of \forestflow \cite{ChavesMontero2025}, an emulator that provides a unified description of \lyaf clustering, now enables a direct connection between them. In this section, we first describe \forestflow, and then we use it to infer the range of three-dimensional clustering models compatible with the DESI DR1 \poned measurements \cite{ChavesMontero2026}. Finally, we validate the resulting predictions using a high-resolution hydrodynamical simulation.

%%%%%%%%%%%%%
%%%%%%%%%%%%%

\subsection{ForestFlow}
\label{sec:mapping_forestflow}

To provide an unified description of \lyaf clustering, \forestflow maps cosmological and IGM parameters onto the large-scale \lya bias parameters, $b_\delta$ and $\beta$, as well as a set of parameters describing deviations of the three-dimensional clustering from linear theory. This distinguishes \forestflow from the \texttt{lace-mpg} emulator used in the DESI DR1 \poned analysis (Section~\ref{sec:data_p1d}): while \forestflow predicts the three-dimensional clustering over a broad range of scales, \texttt{lace-mpg} is designed to model one-dimensional clustering from intermediate to small scales. Below, we briefly summarize the \forestflow framework and refer the reader to \citet{ChavesMontero2025} for a detailed description.

The large-scale \lya bias parameters can be combined with the linear matter power spectrum to model the three-dimensional \lya flux power spectrum \citep[][]{McDonald2003},
\begin{equation}
    \label{eq:p3d_model}
    \pthreed(k, \mu) = b_\delta^2 (1 + \beta \, \mu^2)^2 P_\mathrm{lin}(k)\, D_\mathrm{NL}(k,\mu),
\end{equation}
where $P_\mathrm{lin}$ is the linear matter power spectrum and
\begin{equation}
    \label{eq:dnl}
    D_\mathrm{NL} = \exp \left\{\left[q_1 \Delta^2(k) + q_2 \Delta^4(k) \right] \left[1-\left(\frac{k}{k_\mathrm{v}}\right)^{a_\mathrm{v}} \mu^{b_\mathrm{v}}\right] - \left(\frac{k}{k_\mathrm{p}}\right)^2 \right\},
\end{equation}
is a phenomenological correction that accounts for nonlinear gravitational evolution, small-scale redshift-space distortions, thermal broadening, and pressure smoothing \citep{Arinyo2015}. Here, $\Delta^2(k)\equiv k^3 P_\mathrm{lin}(k)/(2\pi^2)$ denotes the dimensionless linear matter power spectrum, while $q_1$, $q_2$, $k_\mathrm{v}$, $a_\mathrm{v}$, $b_\mathrm{v}$, and $k_\mathrm{p}$ are free parameters.

BAO analyses model three-dimensional clustering through the Fourier transform of Eq.~\ref{eq:p3d_model}, typically setting $D_\mathrm{NL}=1$ because nonlinear corrections affect scales much smaller than the BAO feature. In contrast, accurate modeling of \poned requires the full nonlinear description. We can obtain this observable by integrating \pthreed over transverse modes,
\begin{equation}
    \label{eq:p1d_int}
    P_\mathrm{1D}(k_\parallel)
    = \frac{1}{2\pi}
    \int_{0}^{\infty} \mathrm{d}k_\perp \, k_\perp \,
    P_\mathrm{3D}(k_\parallel, k_\perp),
\end{equation}
where $k_\parallel$ and $k_\perp$ denote the line-of-sight and transverse wavenumbers, respectively, and $k^2 = k_\parallel^2 + k_\perp^2$. Previous studies have shown that Eq.~\ref{eq:p3d_model} accurately reproduces both the one- and three-dimensional clustering when the nonlinear correction parameters are allowed to vary \citep{Arinyo2015, Givans2022, Chabanier2024, ChavesMontero2025}. 

Using measurements from the suite of \texttt{MP-Gadget} cosmological simulations employed to train the \texttt{lace-mpg} emulator (see Section~\ref{sec:data_p1d}), \forestflow was trained to predict the large-scale Ly$\alpha$ bias parameters and the six small-scale parameters as functions of cosmology and IGM physics. To do so, \forestflow adopts the same parameterization as \texttt{lace-mpg}, namely $\Delta^2_\mathrm{p}$, $n_\mathrm{p}$, $\bar{F}$, $\sigma_\mathrm{T}$, $\gamma$, and $k_\mathrm{F}$. The emulator is based on conditional normalizing flows \citep{Winkler2019, Papamakarios2019} and reproduces \pthreed to within $\simeq3\%$ up to $k = 5\,\iMpc$ and \poned to within $\simeq1.5\%$ up to $k_\parallel = 4\,\iMpc$.

Although \forestflow and \texttt{lace-mpg} are trained on the same simulation suite and share the same input parameterization, they employ different architectures and were developed independently. We therefore assess the consistency of their predictions by evaluating both emulators on 10\,000 randomly selected samples from the MCMC chain of the DESI DR1 \poned analysis. The ratio of their \poned predictions has a mean of 0.9979 and a standard deviation of 1.58\%, demonstrating excellent agreement between the two models. Given this level of agreement, and considering the technical complexity of propagating an additional emulator uncertainty through the full analysis pipeline described in the next section, we do not include an extra uncertainty term beyond that of the \texttt{lace-mpg} emulator. The latter is naturally inherited by our mapping from one- to three-dimensional measurements since it is propagated in the DESI \poned analysis.

%%%%%%%%%%%%%%%%%%%%%%%%%%%%%%%%%%%%%%%%%%%%%%%%
%%%%%%%%%%%%%%%%%%%%%%%%%%%%%%%%%%%%%%%%%%%%%%%%

\subsection{\lya forest holography}
\label{sec:mapping_3dfrom1d}

In this section, we use \forestflow to determine the range of three-dimensional clustering models consistent with the constraints from the DESI DR1 \poned analysis \citep{ChavesMontero2026}. Specifically, we evaluate \forestflow on 10\,000 randomly selected samples from the MCMC chains of the \poned analysis (see Section~\ref{sec:data_p1d}) to infer the large-scale Ly$\alpha$ bias parameters together with the six parameters describing the small-scale behavior of \pthreed (Eq.~\ref{eq:dnl}).

\begin{figure}
    \centering
    \includegraphics[width=\linewidth]{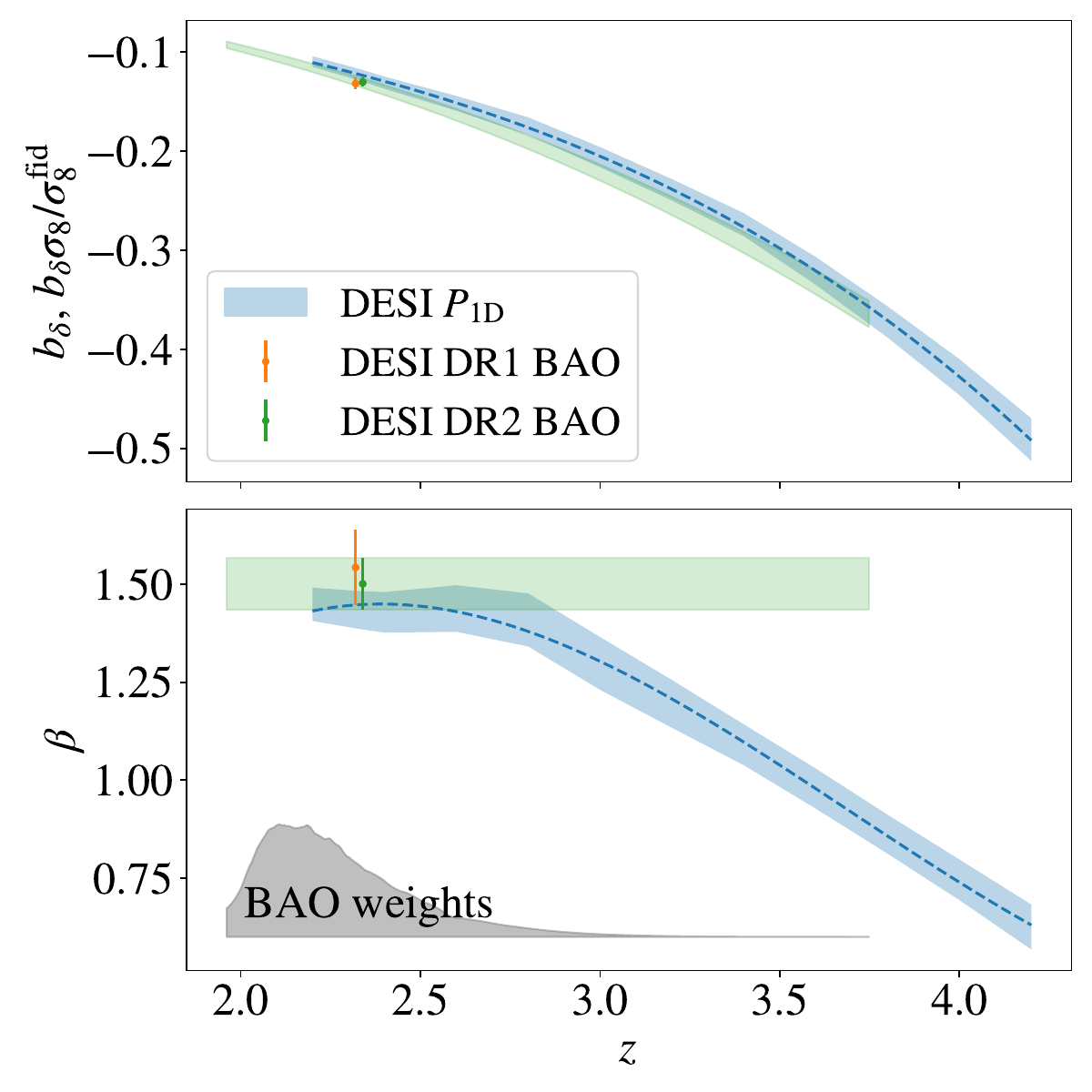}
    \caption{
    Large-scale \lya bias parameters from one- and three-dimensional analyses. The blue shaded regions show constraints on $b_\delta$ and $\beta$ obtained by propagating the DESI \poned measurements through \forestflow, while the orange and green points correspond to $b_\delta \,\sigma_8/\sigma_8^\mathrm{fid}$ and $\beta$ measurements from our DESI DR1 and DR2 BAO reanalyses, respectively. The blue dashed curves show the best-fitting third-order polynomials to one-dimensional measurements, while the green shaded regions indicate the redshift evolution assumed in the DR2 BAO analysis. The gray shaded area shows the redshift distribution of the pairs contributing to DR2 BAO measurements. Error bars and shaded regions represent $1\sigma$ confidence intervals. For visual clarity, the DR1 and DR2 BAO measurements have been shifted by $\Delta z = +0.01$ and $-0.01$, respectively.
    }
    \label{fig:mapping_biases}
\end{figure}

In Fig.~\ref{fig:mapping_biases}, we compare constraints on $b_\delta$ and $\beta$ inferred through this procedure with those on $b_\delta\,\sigma_8/\sigma_8^\mathrm{fid}$ and $\beta$ from the DESI DR1 and DR2 BAO analyses. If the Universe were described by the \textit{Planck} 2018 cosmology, the two sets of constraints would be expected to agree. The \poned constraints propagated through \forestflow are consistent with the BAO measurements at the $1\sigma$ level, indicating that, on large scales, the range of three-dimensional clustering models compatible with the \poned data is also consistent with direct three-dimensional constraints from BAO analyses. We further find that the BAO and \poned analyses yield comparable constraining power on these parameters, despite probing different scales and relying on distinct modelling approaches.

%%%%%%%%%%%%%%%%%%%%%%%%
%%%%%%%%%%%%%%%%%%%%%%%%

The blue dashed lines show the best-fitting third-order polynomials describing the redshift evolution of $b_\delta$ and $\beta$ inferred the \poned analysis, with the corresponding coefficients listed in Table~\ref{tab:p3d_params}. The predicted evolution is in good agreement with the parameterization adopted in BAO analyses, namely $b_\delta(z) = b_\delta(z_\mathrm{eff})[(1+z)/(1+z_\mathrm{eff})]^a$ with $a = 2.9$ \citep{McDonald2006}, and $\beta(z) = \beta(z_\mathrm{eff})$, over the redshift range that carries most of the statistical weight of the BAO measurements, indicated by the gray shaded region. It is also worth noting that the inferred redshift evolution of these parameters is broadly consistent with the predictions of the highest-resolution hydrodynamical simulation in the \accel suite \citep{Chabanier2024}.

%%%%%%%%%%%%%%%%%%%%%%%%
\begin{table}
\caption{Best-fitting coefficients capturing the redshift evolution of the parameters describing three-dimensional clustering (see Eqs.~\ref{eq:p3d_model} and \ref{eq:dnl}), $\theta=a_0 + a_1 x + a_2 x^2 + a_3 x^3$, where $\theta$ runs over all the parameters, $x=(1 + z)/(1+z_0)$, and $z_0=3$.}
\label{tab:p3d_params}
\centering
\begin{tabular}{lrrrr}
\hline
Parameter       & $a_0$ & $a_1$ & $a_2$ & $a_3$ \\
\hline
$b_\delta$      & 0.29 & -1.13 & 1.40 & -0.76 \\
$\beta$         & -8.59 & 29.19 & -26.96 & 7.66 \\
$q_1$           & 7.11 & -22.12 & 22.88 & -7.43 \\
$q_2$           & -12.47 & 39.48 & -40.22 & 13.43 \\
$a_\mathrm{v}$  & 15.00 & -42.99 & 41.32 & -12.91 \\
$b_\mathrm{v}$  & 6.70 & -15.54 & 15.61 & -5.05 \\
$k_\mathrm{v}\,[\mathrm{Mpc}^{-1}]$ & 14.45 & -40.06 & 35.71 & -9.77 \\
$k_\mathrm{p}\,[\mathrm{Mpc}^{-1}]$ & -244.13 & 738.77 & -713.87 & 233.83 \\
\hline
\end{tabular}
\end{table}
%%%%%%%%%%%%%%%%%%%%%%%%

Furthermore, we compare our results with direct constraints on $b_\delta$ obtained from the analysis of the DESI DR1 \poned measurements using a model closely related to Eqs.~\ref{eq:p3d_model} and \ref{eq:dnl} \citep{Karacayli2025_p1d_dr1}. To facilitate this comparison, we fit the inferred redshift evolution of $b_\delta$ using the same power-law parameterization adopted in BAO analyses. We obtain $b_\delta(z_\mathrm{eff})=-0.2111$ and $a=3.073$, in reasonable agreement with the values inferred from the direct analysis, $b_\delta(z_\mathrm{eff})=-0.2175 \pm 0.0018$ and $a=2.963 \pm 0.058$. The observed offsets are likely driven by differences in the small-scale parameterization, the input datasets, and the treatment of systematic effects adopted in the two analyses.

\begin{figure}
    \centering
    \includegraphics[width=\linewidth]{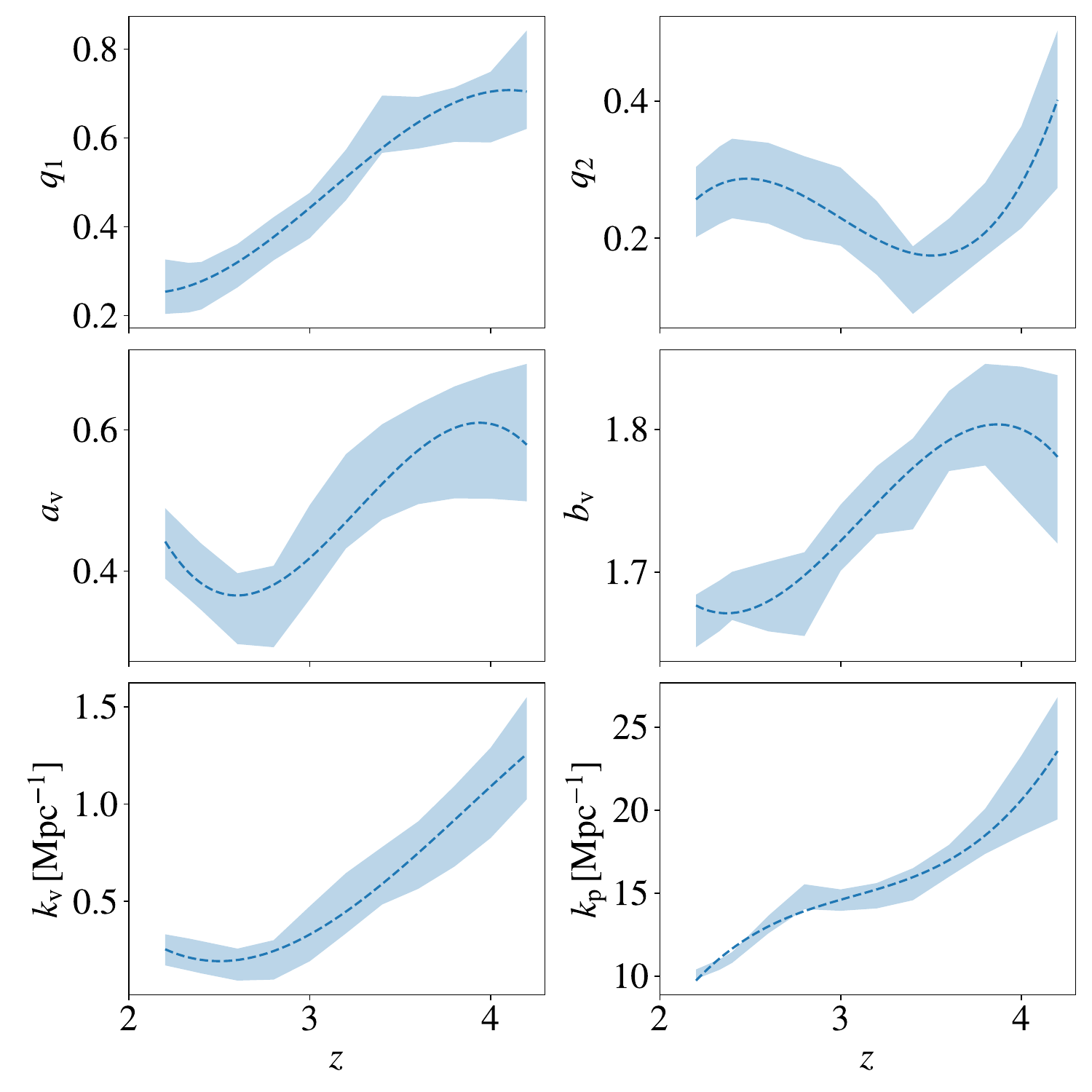}
    \caption{
        Constraints on the parameters describing the small-scale behavior of \pthreed (see Eq.~\ref{eq:dnl}) inferred from the DESI \poned analysis using \forestflow. The dashed lines show the best-fitting third-order polynomial model for the redshift evolution of each parameter, and the shaded regions indicate the $1\,\sigma$ confidence intervals.
    }
    \label{fig:dnl}
\end{figure}

In Fig.~\ref{fig:dnl}, we show the redshift evolution of the parameters that describe departures of \pthreed from linear theory on small scales (see Eq.~\ref{eq:dnl}). The dashed lines indicate the best-fitting third-order model for the redshift evolution of each parameter, and the corresponding coefficients are listed in Table~\ref{tab:p3d_params}. The inferred trends are broadly consistent with those found in hydrodynamical simulations \citep{Arinyo2015, Chabanier2024}. However, comparisons with simulation results should be interpreted with some caution, as several of these parameters have partially degenerate effects on \pthreed. We use these constraints are priors in a companion full-shape analysis of DESI DR2 measurements \citep{DESIDR2full}.

%%%%%%%%%%%%%%%%%%%%%%%%
%%%%%%%%%%%%%%%%%%%%%%%%

%%%%%%%%%%%%%%%%%%%%%%%%%%%%%%%%%%%%%%%%%%%%%%%%
%%%%%%%%%%%%%%%%%%%%%%%%%%%%%%%%%%%%%%%%%%%%%%%%

\subsection{Validation}
\label{sec:mapping_validation}

In the previous section, we used \forestflow to translate constraints from the DESI \poned analysis into predictions for three-dimensional \lya clustering. Although the inferred large-scale clustering parameters are in excellent agreement with direct BAO measurements, both the \poned analysis and the \forestflow emulator rely on a suite of hydrodynamical simulations with moderate resolution. While \citet{ChavesMontero2026} showed that the cosmological constraints derived from the DESI \poned measurements are largely insensitive to effects derived from the limited resolution of these simulations, it remains important to verify that the mapping from one- to three-dimensional clustering is similarly robust.

To perform this test, we use one of the simulations from the \accel project, a suite of six cosmological hydrodynamical simulations run with the adaptive mesh code \texttt{Nyx} \citep{Almgren2013, Sexton2021}. We employ the highest-resolution simulation in the suite, which was run assuming a {\it Planck} 2016 cosmology \citep{Planck2016} and has a box size of 237~Mpc with $6144^3$ resolution elements, corresponding to a physical resolution of 37~kpc. This resolution is sufficient to achieve percent-level accuracy in \lyaf predictions throughout the range of scales and redshifts relevant for DESI \citep{Bolton2009, Lukic2015}. Additional details are provided by \citet{Chabanier2024}.

First, we construct a DESI-like \poned mock based on predictions from the \accel simulation. We begin by smoothing the \poned measurements from \accel to mitigate the impact of cosmic variance arising from the finite simulation volume. Following \citet{ChavesMontero2026}, we model the smoothed \poned as the product of the \forestflow prediction for the \texttt{mpg-central} simulation and a five-parameter smooth function. The former provides the baseline template, while the parameters of the latter are optimized to capture the residual dependence on cosmology and IGM physics. We then interpolate the smoothed predictions to the redshifts and scales of the DESI DESI \poned measurements, and analyze the resulting mock with \cupid using the same covariance matrix adopted in the DESI DR1 analysis. This procedure yields constraints on the amplitude and slope of the linear matter power spectrum, $\Delta_\mathrm{p}^2$ and $n_\mathrm{p}$, together with the IGM parameters $\bar F$, $\sigma_\mathrm{T}$, $\gamma$, and $k_\mathrm{F}$, in the same way as for the observational data. We subsequently propagate these constraints through \forestflow following the procedure described in the previous section. Further details of the mock construction and fitting methodology can be found in \citet{ChavesMontero2026}.

\begin{figure}
    \centering
    \includegraphics[width=\linewidth]{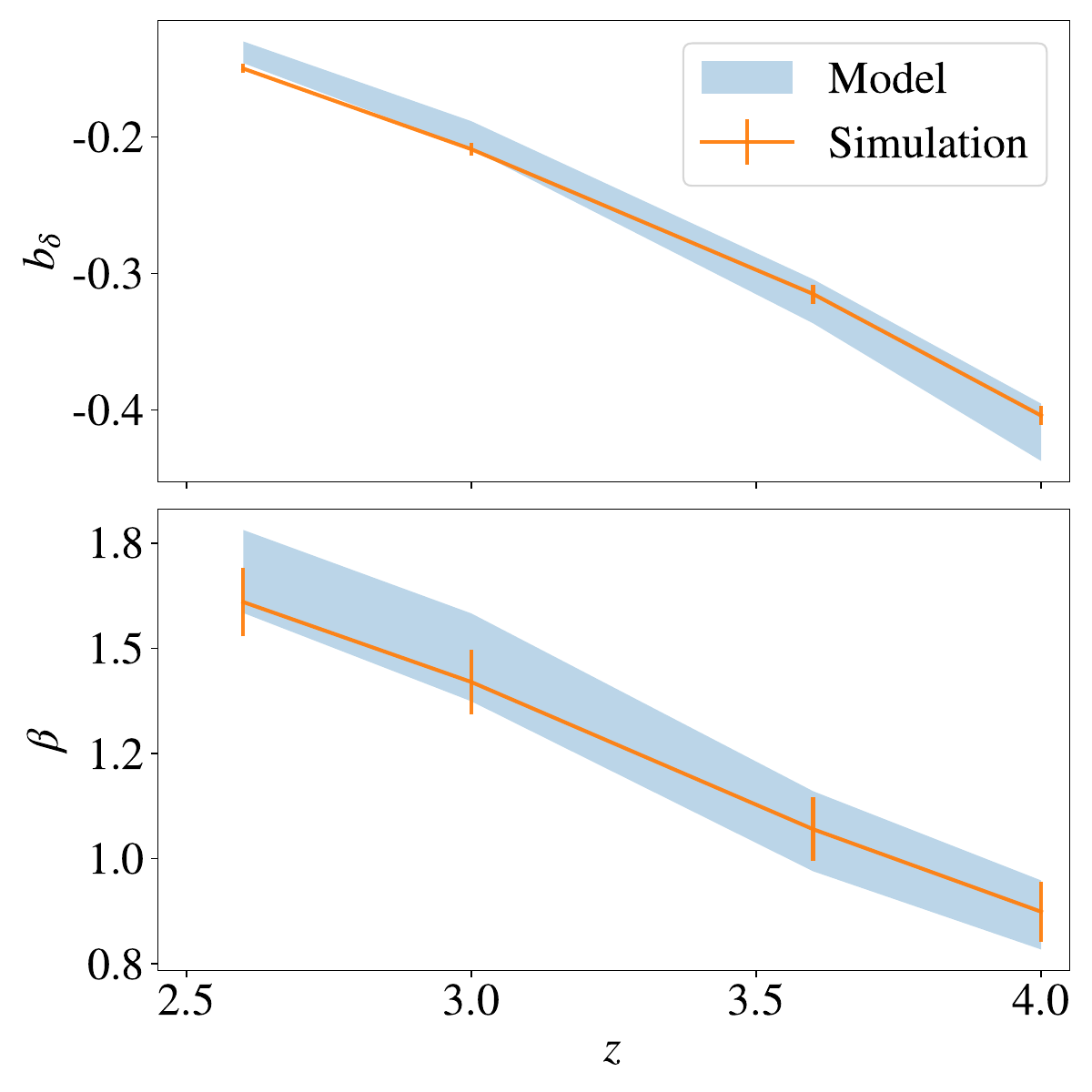}
    \caption{
    Validation of the large-scale three-dimensional clustering predictions produced by our methodology. The blue shaded regions show the predictions for $b_\delta$ and $\beta$ inferred from the analysis of a DESI-like \poned mock based on the \accel simulation, while the orange lines show direct measurements of these parameters from the \accel simulation.
    }
    \label{fig:bias_accel2}
\end{figure}

In Fig.~\ref{fig:bias_accel2}, we compare the inferred constraints on $b_\delta$ and $\beta$ with the best-fitting values obtained from a direct fit to the \accel \pthreed measurements using Eq.~\ref{eq:p3d_model} \citep[see][]{Chabanier2024}. The shaded regions indicate the standard deviation of the inferred parameters from the MCMC chain, while the error bars show the uncertainties from the direct fit. Despite the fact that the \poned mock is fitted and subsequently mapped to three-dimensional clustering using simulations generated with a different numerical code and at a different resolution than those of \accel, we find excellent agreement between the inferred and directly fitted bias parameters.

%%%%%%%%%%%%%%%%%%%%%%%%%%%%%%%%%%%%%%%%%%%%%%%%
%%%%%%%%%%%%%%%%%%%%%%%%%%%%%%%%%%%%%%%%%%%%%%%%

\begin{figure}
    \centering
    \includegraphics[width=\linewidth]{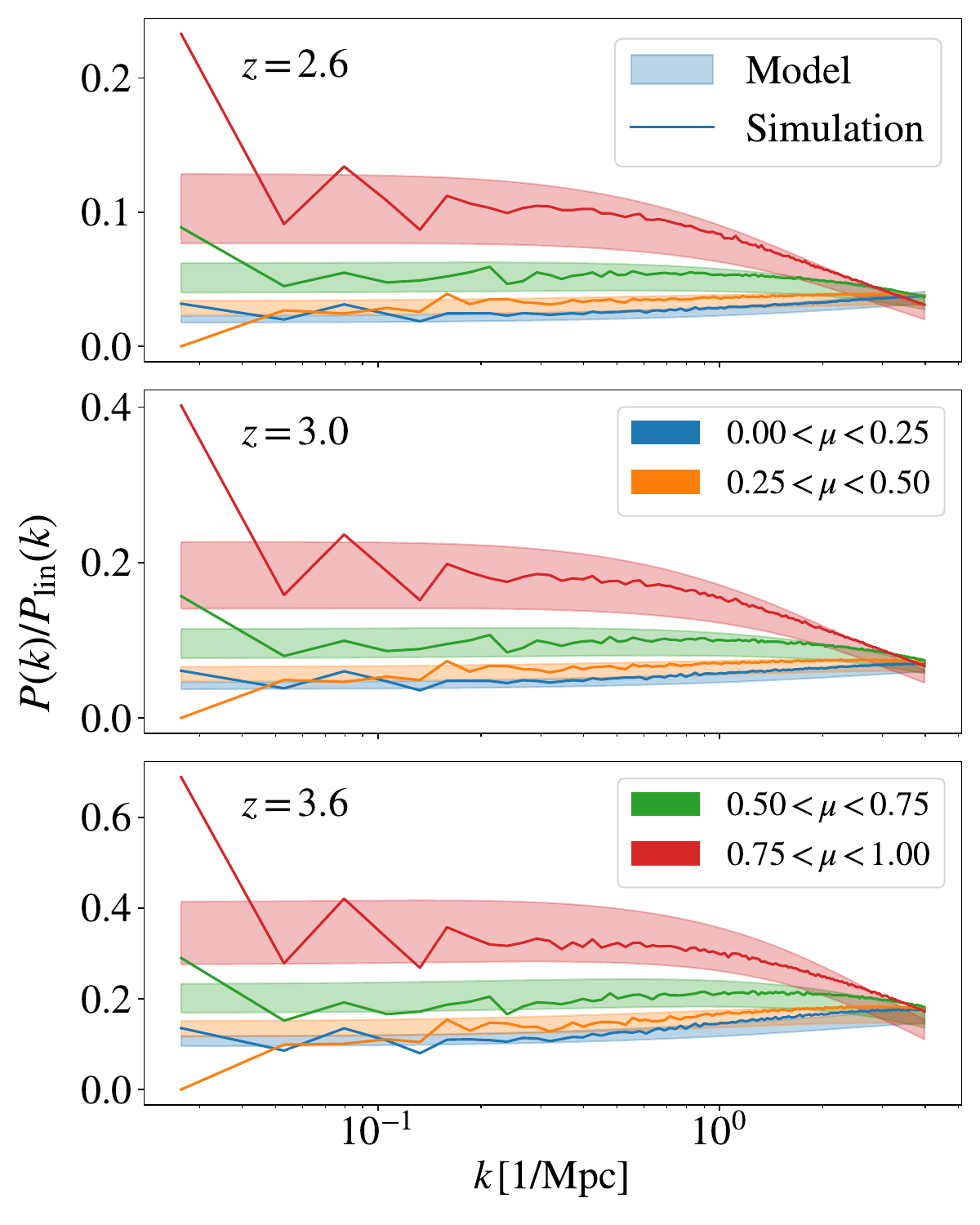}
    \caption{
    Validation of the three-dimensional clustering predictions produced by our methodology. The shaded regions show predictions inferred from the analysis of a DESI-like \poned mock based on the \accel simulation, while the lines correspond to direct \pthreed measurements from the \accel simulation. The top, middle, and bottom panels show results at $z=2.6$, 3.0, and 3.6, respectively. Colors denote different orientations relative to the line of sight, with $\mu=0$ and 1 corresponding to the perpendicular and parallel directions, respectively.
    }
    \label{fig:p3d_accel2}
\end{figure}

In Fig.~\ref{fig:p3d_accel2}, we assess the accuracy of our methodology in recovering three-dimensional clustering constraints across the full range of scales relevant for DESI by comparing our predictions with direct measurements of \pthreed from \accel. We generate these predictions by evaluating Eq.~\ref{eq:p3d_model} over 10\,000 random samples drawn from the best-fitting MCMC chain to the \accel mock, with shaded regions indicating the corresponding $1\sigma$ intervals. We find excellent agreement between the predicted and measured power spectra across all redshifts and scales shown, including the smallest scales accessible to DESI, where potential resolution-dependent systematics are expected to be most significant.

%%%%%%%%%%%%%%%%%%%%%%%%%%%%%%%%%%%%%%%%%%%%%%%%
%%%%%%%%%%%%%%%%%%%%%%%%%%%%%%%%%%%%%%%%%%%%%%%%

Taken together, these findings show that the one-to-three-dimensional mapping provided by \forestflow accurately recovers both the large-scale bias parameters and the small-scale clustering measured in a high-resolution hydrodynamical simulation. This exercise thus validates our methodology across the full range of scales relevant for DESI \lyaf analyses.
\section{Combining one- and three-dimensional analyses}
\label{sec:combined}

The framework described in the previous section provides a direct mapping between one- and three-dimensional Lyman-$\alpha$ forest clustering and could, in principle, be used to perform a joint analysis of \poned and BAO data. At present, however, such an analysis is limited by the lack of a fully consistent treatment of contaminants across the two datasets. These differences could lead to inconsistencies between the corresponding one- and three-dimensional clustering predictions, while also introducing unnecessary parameter redundancy. Instead, to forecast the potential gains of a fully joint analysis, we proceed to combine the posterior distributions obtained independently from the \poned and BAO analyses.

\begin{figure}
    \centering
    \includegraphics[width=\linewidth]{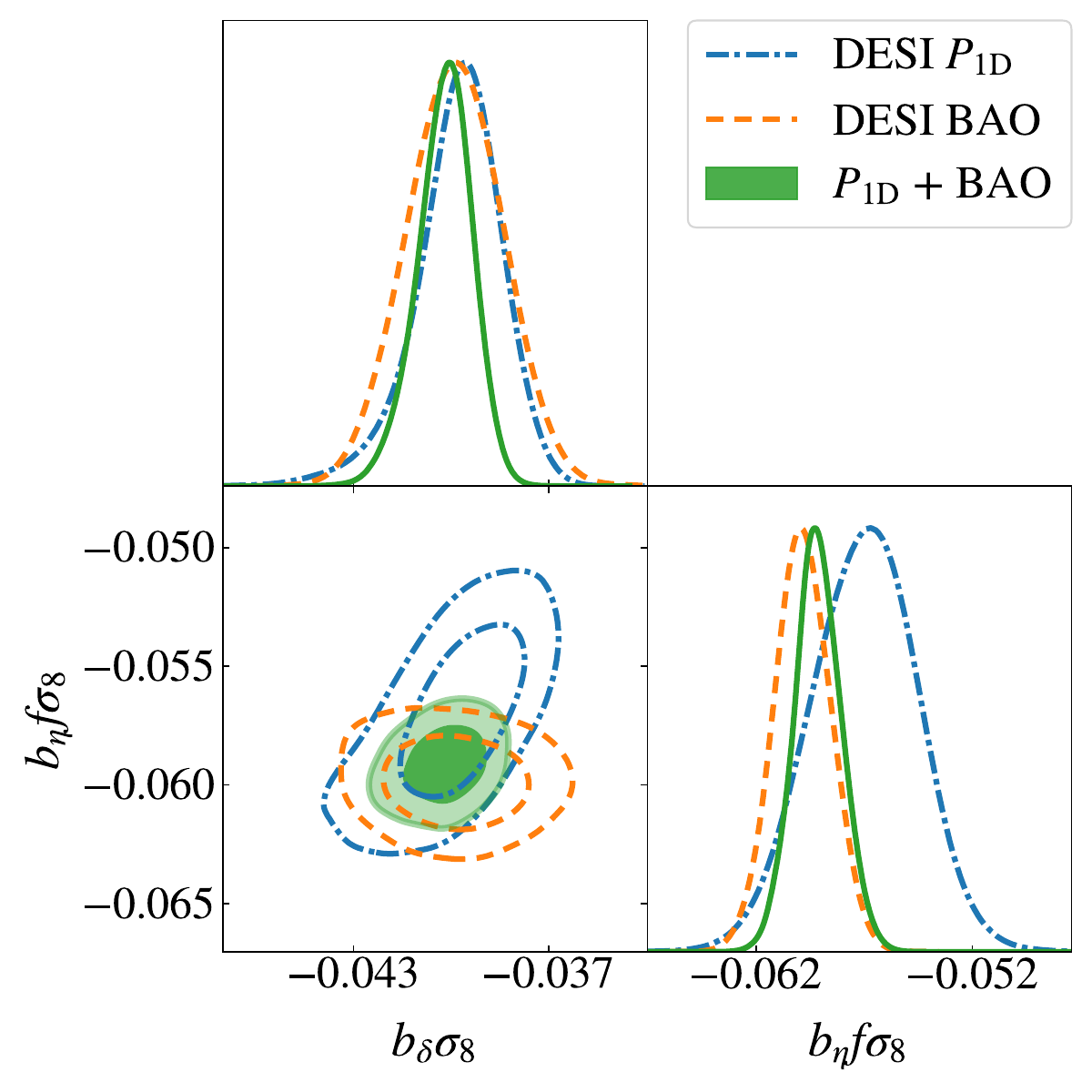}
    \caption{
    Constraints on $b_\delta \sigma_8$ and $b_\eta f \sigma_8$ at $\zeff=2.33$. The blue dotted contours show constraints obtained by propagating the DESI \poned results through \forestflow, while the orange dashed contours correspond to the DESI DR2 BAO analysis. The green shaded region shows the joint constraints obtained by combining both analyses.}
    \label{fig:bdsig8_bfsig8}
\end{figure}

BAO analyses constrain, among other quantities, the parameter combinations $b_\delta \sigma_8$ and $b_\eta f \sigma_8$, with all quantities evaluated at the effective redshift of the analysis (see Section~\ref{sec:data_xi3d}). These same combinations can be obtained by propagating the \poned MCMC chain through \forestflow. The only additional step besides the process described in the previous section is that, for each sample in the chain, we compute $\sigma_8$ and $f\sigma_8$ at the effective redshift of BAO measurements using the Boltzmann solver \texttt{camb} \citep{Lewis2000}. In Fig.~\ref{fig:bdsig8_bfsig8}, we compare the resulting constraints from the DESI \poned analysis and the DESI DR2 BAO analysis. The two analyses agree at the $1\,\sigma$ level and achieve comparable precision: the \poned analysis yields slightly tighter constraints on $b_\delta \sigma_8$, while the BAO analysis provides tighter constraints on $b_\eta f \sigma_8$.

%%%%%%%%%%%%%%%%%%%%%%%%%%%%%%%%%%%%%%%%%%%%%%%%%
%%%%%%%%%%%%%%%%%%%%%%%%%%%%%%%%%%%%%%%%%%%%%%%%%

The green shaded region shows the joint constraints obtained by combining the \poned and BAO analyses. We compute the joint posterior via importance sampling, assuming that the \poned and BAO results are uncorrelated. This is a very good approximation because \poned measurements rely on the auto-correlation of individual \lya sightlines, whereas BAO analyses use cross-correlations between different \lya sightlines and between \lya sightlines and quasar positions. See Appendix~\ref{app:correlation} for a quantitative estimation. As can be seen, the joint constraints are significantly tighter than those from either probe alone. This gain is especially significant because the two analyses exhibit different parameter degeneracy directions, making them highly complementary.
\section{Summary and conclusions}
\label{sec:conclusions}

Traditional analyses of \lyaf clustering treat one- and three-dimensional statistics separately because they probe the matter distribution on very different scales. In this work, we use \forestflow \citep{ChavesMontero2025}, an emulator that provides a unified description of \lya clustering from linear to nonlinear scales, to infer the three-dimensional clustering models compatible with one-dimensional measurements. We refer to this framework as \emph{Lyman-$\alpha$ holography}, by analogy with the reconstruction of higher-dimensional structure from lower-dimensional information. The main results of this work are summarized below.

\begin{itemize}
    \item In Fig.~\ref{fig:mapping_biases}, we present constraints on the large-scale \lya clustering parameters $b_\delta$ and $\beta$ obtained by propagating the DESI DR1 \poned constraints \citep{ChavesMontero2026} through \forestflow. The resulting constraints are in excellent agreement with those obtained from direct analyses of three-dimensional \lyaf clustering in the DESI DR1 and DR2 BAO analyses \citep{DESI2024.IV.KP6, DESI.DR2.BAO.lya}.

    \item In Fig.~\ref{fig:dnl}, we show constraints from the DESI \poned measurements on a parametric model describing small-scale deviations of three-dimensional \lya clustering from linear theory. These constraints are adopted as priors in a companion work presenting the full-shape analysis of DESI DR2 measurements \citep{DESIDR2full}.
    
    \item In Section~\ref{sec:mapping_validation}, we validate our methodology using \accel, a large-volume, high-resolution hydrodynamical simulation. As shown in Figs.~\ref{fig:bias_accel2} and \ref{fig:p3d_accel2}, we find excellent agreement between our methodology and direct measurements from \accel across the full range of scales relevant for DESI \lyaf analyses.

    \item In Fig.~\ref{fig:bdsig8_bfsig8}, we compare the constraints on $b_\delta \sigma_8$ and $b_\eta f \sigma_8$ inferred from the DESI DR1 \poned analysis using our methodology with those obtained from the DESI DR2 BAO analysis. We find that the two probes provide comparable constraining power on these parameter combinations while exhibiting markedly different degeneracy directions, highlighting their strong complementarity.
\end{itemize}

This work paves the way toward a fully self-consistent joint cosmological analysis of one- and three-dimensional \lyaf clustering. Our results demonstrate the strong complementarity of these observables, showing that their combination can deliver substantially tighter constraints than either probe alone. The principal remaining challenge is the development of a unified treatment of contaminants that can be applied consistently to both analyses, particularly metal absorption and HCD systems. Overcoming this limitation will enable the framework introduced here to jointly infer cosmological and astrophysical parameters from the full information content of Lyman-$\alpha$ forest clustering.
\begin{acknowledgements}

We thank Calum Gordon and Naim Kara\c{c}ayl{\i} for insightful comments and suggestions. JCM and AFR acknowledge financial support from the Spanish Ministry of Science and Innovation through the PID2024-159420NB-C41 project and the ``Excelencia Severo Ochoa'' program (CEX2024-001441-S from MICIU AEI 10.13039/501100011033) and the European Union through the ERC Consolidator Grant program (COSMO-LYA, grant agreement 101044612). Views and opinions expressed are however those of the authors only and do not necessarily reflect those of the European Union or the European Research Council Executive Agency. Neither the European Union nor the granting authority can be held responsible for them. IFAE is partially funded by the CERCA program of the Generalitat de Catalunya. This research used resources of the National Energy Research Scientific Computing Center (NERSC), a Department of Energy User Facility. 

This material is based upon work supported by the U.S. Department of Energy (DOE), Office of Science, Office of High-Energy Physics, under Contract No. DE–AC02–05CH11231, and by the National Energy Research Scientific Computing Center, a DOE Office of Science User Facility under the same contract. Additional support for DESI was provided by the U.S. National Science Foundation (NSF), Division of Astronomical Sciences under Contract No. AST-0950945 to the NSF’s National Optical-Infrared Astronomy Research Laboratory; the Science and Technology Facilities Council of the United Kingdom; the Gordon and Betty Moore Foundation; the Heising-Simons Foundation; the French Alternative Energies and Atomic Energy Commission (CEA); the National Council of Humanities, Science and Technology of Mexico (CONAHCYT); the Ministry of Science, Innovation and Universities of Spain (MICIU/AEI/10.13039/501100011033), and by the DESI Member Institutions: \url{https://www.desi.lbl.gov/collaborating-institutions}. Any opinions, findings, and conclusions or recommendations expressed in this material are those of the author(s) and do not necessarily reflect the views of the U. S. National Science Foundation, the U. S. Department of Energy, or any of the listed funding agencies.

The authors are honored to be permitted to conduct scientific research on I'oligam Du'ag (Kitt Peak), a mountain with particular significance to the Tohono O’odham Nation.

\end{acknowledgements}
\section*{Data availability}

The \forestflow\footnote{\url{https://github.com/igmhub/ForestFlow}} and \texttt{lace-mpg}\footnote{\url{https://github.com/igmhub/LaCE}} emulators are publicly available, together with the \texttt{cup1d}\footnote{\url{https://github.com/igmhub/cup1d}} likelihood code used in the DESI DR1 \poned analysis \citep{ChavesMontero2026} and the \texttt{vega}\footnote{\url{https://github.com/andreicuceu/vega}} package employed in our reanalysis of the BAO measurements. The data points shown in the figures will be released in machine-readable form on Zenodo, while the notebooks used to produce the figures are available in the \forestflow repository\footnote{\url{https://github.com/igmhub/forestflow}}. 

We acknowledge that this work made direct use of the following \texttt{python} packages: 
\texttt{astropy} \citep{astropy:2013, astropy:2018, astropy:2022},
\texttt{camb} \citep{Lewis2000, Howlett2012_camb}, 
\texttt{emcee} \citep{foremanmackey13}, 
\texttt{freia} \citep{freia},
\texttt{getdist} \citep{Lewis:2019},
\texttt{matplotlib} \citep{matplotlib}, 
\texttt{mpi4py} \citep{dalcin2005_mpi, dalcin2008_mpi, dalcin2011_mpi, Dalcin2021_mpi, Rogowski2023_mpi}, 
\texttt{numpy} \citep{Harris:2020}, 
\texttt{pytorch} \citep{pytorch2019}, 
\texttt{scipy} \citep{virtanen2020_SciPyFundamentalalgorithms}, and
\texttt{scikit-learn} \cite{Pedregosa2011_scikitlearn}.

\bibliographystyle{aa_url}
\bibliography{main, desi}

\begin{appendix}
\section{Reanalysis of DESI BAO measurements}
\label{app:hcd_prior}

Throughout this work, we compare predictions for the large-scale \lya bias parameters derived from the DESI DR1 \poned analysis \citep{ChavesMontero2026} with direct constraints from the DESI DR1 and DR2 BAO measurements \citep{DESI2024.IV.KP6, DESI.DR2.BAO.lya}. To ensure a consistent comparison, we reanalyze the DESI DR1 and DR2 BAO measurements using a common fitting range: $r > 30\,\hMpc$. This differs from the fiducial DR1 analysis, which included scales down to $r=10\,\hMpc$. We further use subsamples with reduced residual DLA contamination and impose a physically motivated prior on the remaining HCD contribution. We describe the reanalysis below, and compare the results with the fiducial ones.

To model three-dimensional \lya clustering, BAO analyses include several contaminants, such as metal absorption and HCD systems. While these have a negligible impact on the inferred BAO scale, they can significantly affect the inferred large-scale \lya bias parameters. Among them, HCD systems are particularly important, as their broad absorption profiles distort correlations along the line of sight \citep{McQuinn2011, FontRibera2012a, Rogers2018_3DHCDs, Tan2025}. On sufficiently large scales, the effect of HCDs can be absorbed into effective bias parameters \citep{FontRibera2012a, Bautista2017},
\begin{align}
\label{eq:lya_hcd}
\tilde{b}_\delta &= b_\delta + b_\mathrm{HCD} F_\mathrm{V}(k_\parallel),\\
\tilde{b}_\delta \, \tilde{\beta} &= b_\delta \, \beta + b_\mathrm{HCD} \, \beta_\mathrm{HCD}\, F_\mathrm{V}(k_\parallel),
\end{align}
where $b_\mathrm{HCD}$ and $\beta_\mathrm{HCD}$ denote the bias and redshift-space distortion parameters associated with HCD absorption, respectively, and $F_\mathrm{V}$ encodes the impact of the column density distribution,
\begin{equation}
    F_\mathrm{V}(k_\parallel) = \frac{\int \mathrm{d}n\, f(n)\, b^\mathrm{halo}_\mathrm{HCD}(n) \, W(k_\parallel, n)}{\int \mathrm{d}n\, f(n)\, b^\mathrm{halo}_\mathrm{HCD}(n) \, W(n)}.
\end{equation}
Here, $n = \log N_\mathrm{HI}$, $f(n)$ is the normalized column density distribution, $b^\mathrm{halo}_\mathrm{HCD}$ is the bias of the host halos of HCD systems, and $W(k_\parallel, n)$ and $W(n)$ are the Fourier transform and equivalent width of the Voigt profile, respectively. On large scales, and assuming negligible dependence of halo bias on column density, this reduces to $b_\mathrm{HCD} = b^\mathrm{halo}_\mathrm{HCD} (\bar{F}_\mathrm{HCD}^{-1} - 1)$ \citep{Tan2025}, where $\bar{F}_\mathrm{HCD} = 1 - \rho_\mathrm{HCD} \int \mathrm{d}n\, f(n)\, W(n)$ is the mean absorption caused by HCD systems and $\rho_\mathrm{HCD}$ denotes the number density of HCDs per unit of forest length.

The DESI DLA finder identifies HCD absorbers and, prior to the BAO analysis, masks the strongest absorption regions while modeling the corresponding damping wings \citep{Ho2021, Wang2022, Brodzeller2025_desiDLA}. In the ideal case of perfect identification, HCD systems would not contribute to the observed Ly$\alpha$ forest clustering signal. In practice, however, the finite spectral resolution and limited signal-to-noise ratio of DESI spectra reduce the completeness of the DLA catalog. In particular, the algorithm achieves high completeness for systems with $\log(N_{\rm HI}[\mathrm{cm}^{-2}]) > 20.3$ in high signal-to-noise spectra, but its performance deteriorates for lower column densities and noisier data. To minimize the impact of residual DLA contamination, we therefore restrict our reanalyses to forests with signal-to-noise ratios greater than 4.5. This corresponds to one of the data splits used in the DESI DR1 and DR2 BAO analyses, for which the DLA finder recovers the vast majority of DLAs. Moreover, this selection closely matches that adopted in the DESI \poned analysis, leading us to expect comparable levels of residual contamination in the one- and three-dimensional analyses.

On the other hand, the BAO analysis is affected by residual contamination from undetected DLAs and absorbers in the range $17.2 < \log(N_{\rm HI}[\mathrm{cm}^{-2}]) < 20.3$, with column densities too small to be identified by DESI. To mitigate the resulting degeneracy between the large-scale \lya bias parameters and HCD contributions, we impose a prior on $b_\mathrm{HCD}$. We derive this prior using the column density distribution over the previous range from \texttt{pyigm}\footnote{\url{https://github.com/pyigm/pyigm}} \citep{Prochaska2017}, which is calibrated to \citet{Prochaska2014}. This yields $1 - \bar{F}_\mathrm{HCD}^{-1} \simeq 0.008$ at the effective redshift of the BAO measurements. Combining this value with observational constraints on the halo bias of HCD hosts, $b_\mathrm{HCD}^\mathrm{halo}\simeq 2$ \citep{FontRibera2012b, PerezRafols2018, PerezRafols2023}, we obtain $b_\mathrm{HCD}\simeq -0.02$. To account for uncertainties in the column density distribution, residual dependence of the halo bias on column density, and incomplete DLA identification even in high signal-to-noise spectra, we adopt a broader Gaussian prior for the reanalyses, $b_\mathrm{HCD}=-0.020\pm0.005$. This is consistent with the best-fitting value from the fiducial DESI DR2 analysis, $b_\mathrm{HCD}=-0.0206\pm0.0090$ \citep{DESI.DR2.BAO.lya}.

\begin{figure}
    \centering
    \includegraphics[width=\linewidth]{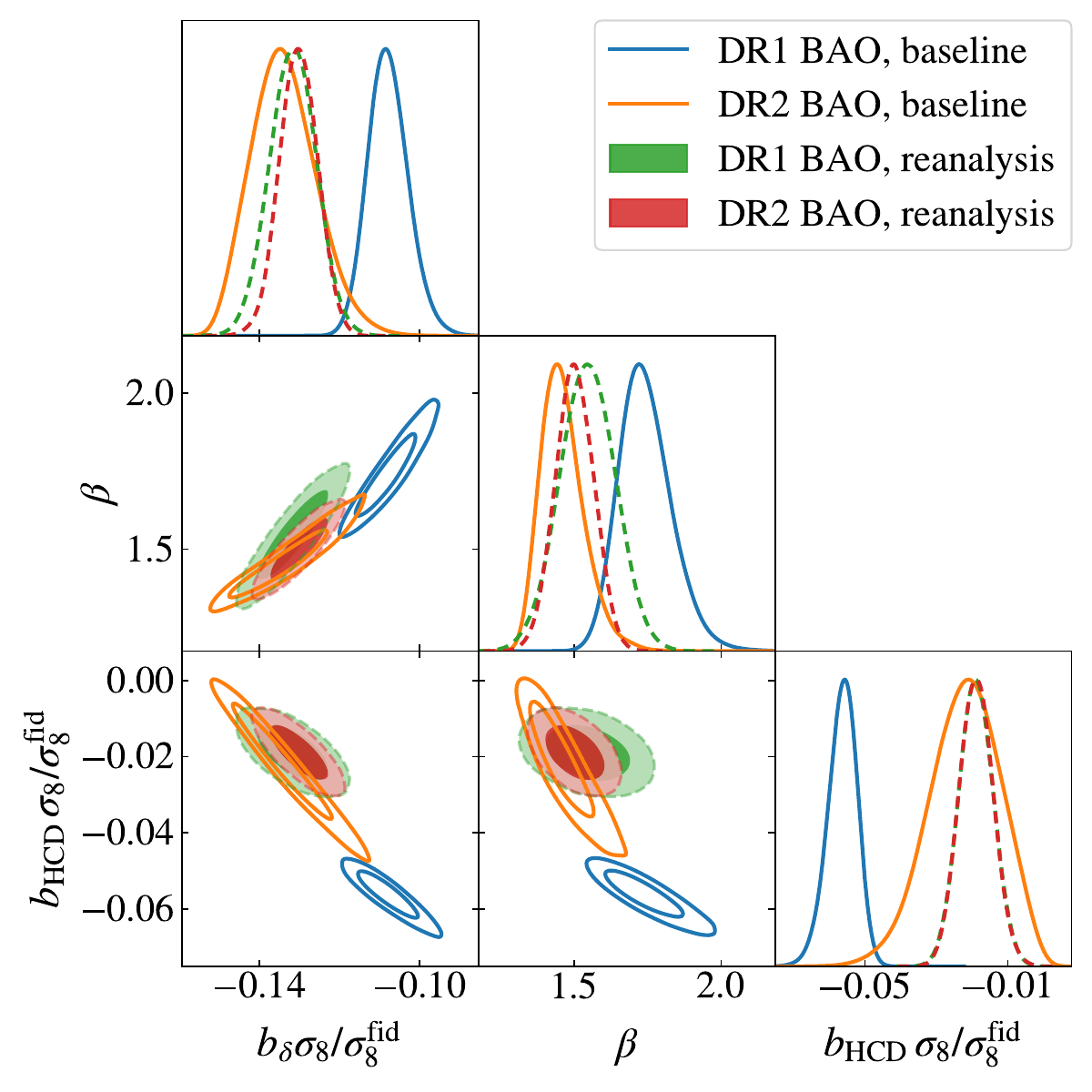}
    \caption{Large-scale \lya bias parameters and level of HCD contamination in DESI BAO analyses. The blue and orange contours correspond to the fiducial DR1 and DR2 analyses, respectively, while the green and red shaded regions show the corresponding reanalyses when using the same range of scales for the fit, only forests from high signal-to-noise spectra, and a prior on $b_{\rm HCD}$.
    }
    \label{fig:hcd_prior}
\end{figure}

Figure~\ref{fig:hcd_prior} compares constraints on the large-scale \lya bias parameters and $b_\mathrm{HCD}$ from the baseline analyses with those from our reanalyses. The DR1 and DR2 baseline results are not fully consistent, with the discrepancy primarily aligned with the degeneracy direction in Eq.~\ref{eq:lya_hcd}. This motivates the reanalysis presented here. In particular, the fiducial DR1 fit prefers values of $b_\mathrm{HCD}$ that are difficult to reconcile with current observational constraints, suggesting that the result is driven by parameter degeneracies rather than physical HCD contamination. In contrast, the reanalysis results are mutually consistent and in excellent agreement with the fiducial DR2 constraints.

\section{Correlation between one- and three-dimensional measurements}
\label{app:correlation}

In Section~\ref{sec:combined}, we combine one- and three-dimensional measurements under the assumption that they are uncorrelated. A precise estimate of their covariance would require mock catalogs, since \poned measurements are based on the autocorrelation along individual \lya sightlines, whereas the three-dimensional BAO analysis relies exclusively on the cross-correlation between different sightlines. Constructing such mocks is beyond the scope of this work. Instead, we present a simple toy model that demonstrates that, even though one- and three-dimensional measurements are sensitive to the same underlying density field, the correlation between these is expected to be small.

To estimate the correlation between the one- and three-dimensional measurements, we first obtain a noiseless \pthreed model by evaluating Eq.~\ref{eq:p3d_model} at the best-fitting parameters of the DESI DR1 \poned analysis at $z=2.33$. We then generate 20\,000 Gaussian realizations of the model by adding fluctuations with a variance proportional to the square of the signal divided by the number of Fourier modes. This approximation therefore includes only the contribution from cosmic variance, neglecting additional sources of uncertainty such as the discrete sampling of the \lya field by quasars, observational noise, and supersample covariance \citep{McDonaldEisenstein2007}. Finally, for each noisy \pthreed realization, we compute the corresponding \poned by integrating over the perpendicular Fourier modes (Eq.~\ref{eq:p1d_int}).

\begin{figure}
    \centering
    \includegraphics[width=\linewidth]{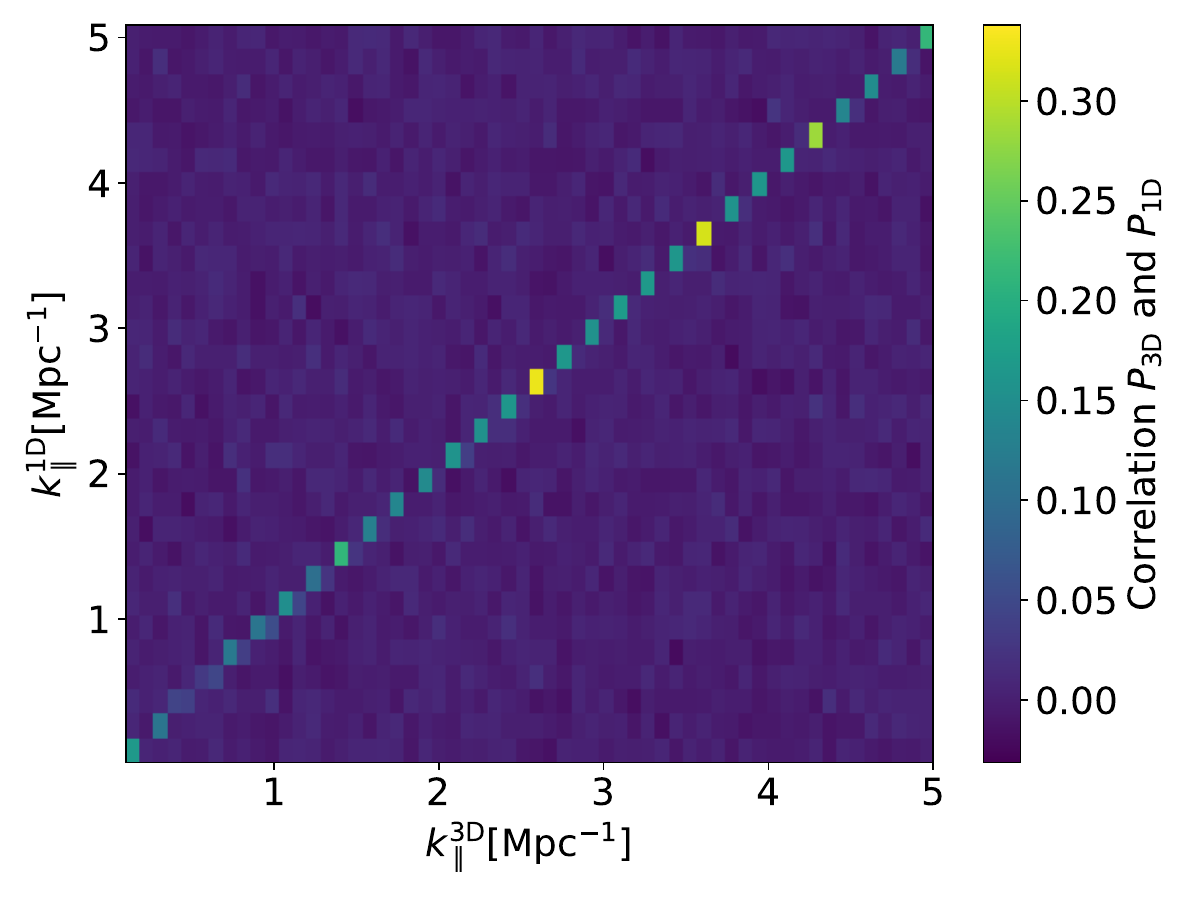}
    \caption{Correlation between the \poned and \pthreed measurements. The x- and y-axes correspond to $k_\parallel^\mathrm{3D}$ and $k_\parallel^\mathrm{1D}$, respectively, while the color scale indicates the correlation coefficient. As expected, the correlation is negligible when $k_\parallel^\mathrm{3D}$ and $k_\parallel^\mathrm{1D}$ differ.}
    \label{fig:corr_p1d_p3d}
\end{figure}

In Fig.~\ref{fig:corr_p1d_p3d}, we show the correlation between the \poned and \pthreed measurements computed from the Gaussian realizations described above. As expected, the correlation is negligible when $k_\parallel^\mathrm{1D}$ and $k_\parallel^\mathrm{3D}$ are different. When the two wavenumbers are comparable, the correlation approximately ranges from zero to one third, reaching a maximum at $k_\parallel^\mathrm{3D}/k_\parallel^\mathrm{1D}\simeq0.66$. We therefore conclude that treating the one- and three-dimensional measurements as uncorrelated is a good approximation for the purposes of this work.

\end{appendix}

\end{document}